# Inhomogeneous critical current in nanowire superconducting single-photon detectors

R. Gaudio[*], K.P.M. op 't Hoog, Z. Zhou, D. Sahin[x], A. Fiore

COBRA Research Institute, Eindhoven University of Technology, P.O. Box 513, NL-5600MB Eindhoven, The Netherlands

Abstract

A superconducting thin film with uniform properties is the key to realize nanowire superconducting single-photon detectors (SSPDs) with high performance and high yield. To investigate the uniformity of NbN films, we introduce and characterize simple detectors consisting of short nanowires with length ranging from 100nm to 15µm. Our nanowires, contrary to meander SSPDs, allow probing the homogeneity of NbN at the nanoscale. Experimental results, endorsed by a microscopic model, show the strongly inhomogeneous nature of NbN films on the sub-100nm scale.

[*]r.gaudio@tue.nl

[x]present address: Centre for Quantum Photonics, H. H. Wills Physics Laboratory, University of Bristol, Tyndall Avenue, Bristol BS8 1TL, UK

Superconducting single-photon detectors (SSPDs)[1,2] based on niobium nitride (NbN) nanowires present high speed, low dark counts and low jitter[3,4]. Compared to more efficient SSPDs recently obtained from amorphous superconducting materials[5-9], NbN SSPDs allow operation at higher temperatures and with simpler read-out circuitry. Due to such favourable combination, SSPDs have brought a breakthrough in fields such as quantum key distribution (QKD)[10,11], nanoscale imaging[12] and quantum optics[13].

However, the number of possible applications is still limited by the low fabrication yield. Indeed, realizing arrays of SSPDs with the same performance is challenging. Understanding and solving this issue could enable free-space single-photon imaging[14], spatial and photon-number resolution[15-17] as well as circumventing dead time limitations in interplanetary optical communication[18].

The physical reason for the poor reproducibility must be sought in the SSPD operating principle: for efficient operation, NbN SSPDs must be biased with a bias current $I_b$ slightly lower than the critical current $I_c$. In this condition, the diffusion of photocreated quasiparticles, together with the vortex unbinding, results in a voltage pulse in the readout circuit[19]. This requires the nanowire to be extremely homogeneous, so that the critical current is uniform along its entire length.

Recent studies[20,21] on meander SSPDs showed a variation in detection efficiency among nominally identical devices. The variation has been ascribed to highly localized areas of the nanowire, named constrictions, characterized by a reduced cross section. The large active area of the meander SSPD used in those studies, though, does not allow a straightforward investigation of such defects. Indeed, the dimensions, density and physical origin of these constrictions are still unknown. In order to address this questions, here we investigate simpler detectors whose active areas consist of short wires with different lengths (L) and fixed width (w=100nm). With these simple devices not only we unequivocally prove that the critical current is not uniform, but we also investigate the typical length scale of its variations. Our experimental results, endorsed by a microscopic model, reveal that the nanowire is not affected by isolated point-like constrictions but is rather continuously inhomogeneous and the typical correlation length of the critical current variation is shorter than 100nm.

The fabrication process starts with the deposition of a 5nm-thick NbN film on a GaAs (001) substrate by means of DC reactive magnetron sputtering. The NbN film is deposited by sputtering a Nb target in Ar+N$_2$ mixture at total pressure $P_{TOT}$=2.3mTorr. The deposition is carried out at a nominal temperature of 400 °C, with target current of 250mA and target voltage of 380V. These deposition conditions, similar to those previously used to fabricate high-performance meander- and waveguide-SSPDs[22-24], allowed the realization of a film with critical temperature $T_c$=9.67K$ and transition width $\Delta T_c$=0.34K. Then contact pads, consisting of 14nm Ti and 140nm Au layers, are defined through optical lithography, electron beam evaporation and lift off. In the final step, the nanowires are patterned from the NbN film by electron-beam lithography and reactive-ion etching in Ar/SF$_6$ plasma. The process was optimized in order to fabricate 100nm wide nanowires with lengths ranging from 100nm to 15μm. In order to prevent latching[25] an additional meander (500nm wide and 573μm long) is defined together with the nanowire to provide a series inductance of 103nH[26]. Figure 1 shows the scanning electron microscope (SEM) images of two nanowires with different lengths.

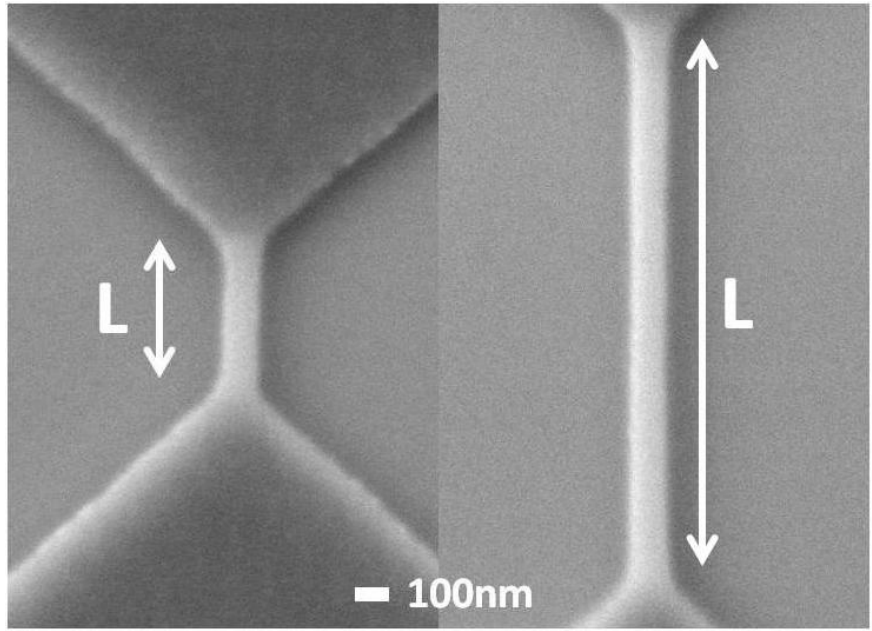

Figure 1. SEM micrographs of a 100nm wide nanowire with length L=400nm (left) and L=1600nm (right).

The electro-optical characterization is performed at an estimated sample temperature of 5K in a cryogenic probe station with an optical window. During the electrical measurements, the detector is biased through the DC port of a bias-T with a voltage source connected in series with a 10Ω bias resistor. The RF port of the bias-T is connected to a 50 Ω cap. The electrical contact with the device is established with a 50 Ω microwave probe connected to the circuit through an SMA coaxial cable.

A chip containing 16 nominally identical nanowires for each length L, with L=100nm, 400nm, 1.6µm and 15µm, is measured in a single cooling run. While it limits the statistics, this procedure ensures that all the devices are characterized at the same effective temperature.

The measured critical currents $I_c$ are displayed in figure 2 as a function of nanowire length. The data, covering more than two orders of magnitude in wire length, shows a clear trend. The critical current, ideally independent of wire length, decreases with increasing length. In addition, the $I_c$ values present a large spread, as measured by the standard deviation $\Delta I_c$, even for the shortest wires, for which the spread is maximum. To highlight the previous observations, we reported the same set of data in the histogram of figure 3 (panel a).

The variation of $I_c$ values among nominally identical devices constitutes a direct proof for the inhomogeneous nature of the wire. A more intriguing feature is the decrease in the ($\bar{I}_c$) encountered already between 100nm and 400nm long nanowires. These observations reveal that each nanowire presents a continuously inhomogeneous distribution of $I_c$ with correlation length shorter than or equal to 100nm. Indeed, if the nanowire was homogeneous on a 400nm length scale, for example, we would not observe any decrease in the average $I_c$ between these and the 100nm long wires. Very similar results were obtained on different deposition batches of NbN/GaAs films at different deposition temperatures. In a separate batch, a decrease in the $\bar{I}_c$ has been observed also between L=0nm (bowtie detector [12]) and L=100nm nanowires, indicating that the correlation length is smaller than 100nm.

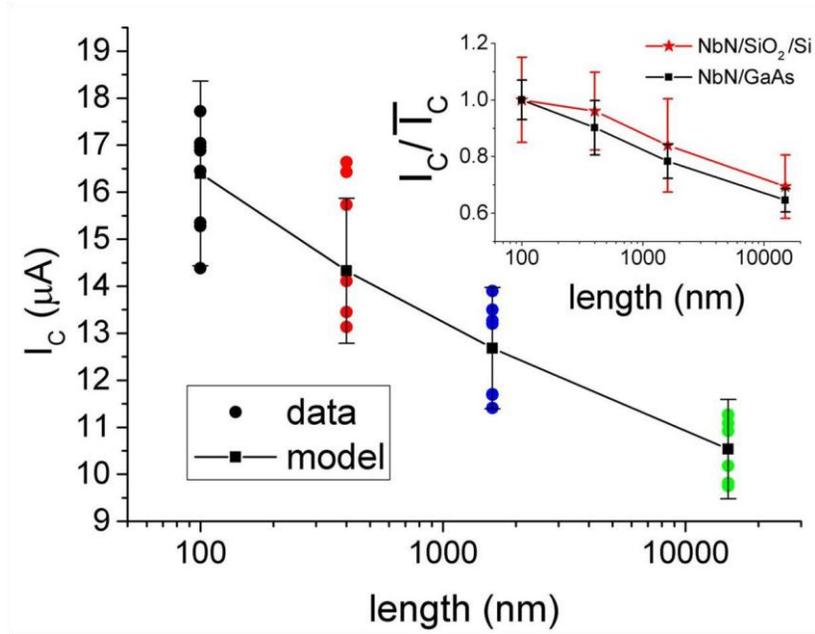

Figure 2. Experimental values of critical currents displayed as a function of nanowire length (filled circles) are superimposed to the $\bar{I}_c$ (black squares) and standard deviation (error bars) calculated with the microscopic model. Inset: The data of the main panel is reported together with the data obtained from a commercially available NbN film. For a clear comparison, both sets of data are normalized to their mean $I_c$ ($\bar{I}_c$) and the standard deviation $\Delta I_c$ is calculated (error bars).

In order to understand whether this behavior is only typical of our NbN films on GaAs, we realized and characterized the same detectors on a commercial NbN film deposited by SCONTEL Superconducting Nanotechnology on a (250nm)SiO$_2$/Si(001) substrate. The NbN film had a critical temperature of $T_c$=10.03K. Since the detectors patterned from this film were measured at a different temperature ~2K, a straightforward comparison with the results of figure 2 (main panel) is only possible with normalized data. In the inset of figure 2 we compare the $\bar{I}_c$ and $\Delta I_c$ values obtained from the two samples where both sets of data are normalized to the $\bar{I}_c$ of their 100nm long wires. The graph clearly shows that the reported behaviours are not unique to our NbN films. In fact, also detectors patterned from the NbN/SiO$_2$ film are characterized by the decrease of $\bar{I}_c$ with increasing length, and they have a large spread in critical current $\Delta I_c$ for each length. It is worth noticing that in both samples the ratio between the normalized $\bar{I}_c$ of the longest and shortest wires is almost 0.6. While we cannot exclude that more homogeneous films may be obtained by varying the deposition conditions, as the results in Ref[19] suggest, the observed inhomogeneity is likely typical of NbN films used for SSPD fabrication.

To gain more insight, we calculated the theoretical depairing current ($I_c^{dep}$) and compared it with the $I_c$ values resulted from our measurements. The depairing current is estimated according to the Ginzburg-Landau model and the temperature dependence proposed by Bardeen[27,28]. In the $I_c^{dep}$ estimation we considered the following quantities: the measured sheet resistance in the normal state $R_s$=870 Ω/sq, the diffusivity D~0.5 cm$^2$/s of NbN films [29], the measured critical temperature $T_c$=9.67K and the energy gap at zero temperature $\Delta_0$=2.07 k$_B$T$_c$ [29]. For a 100nm wide nanowire operating at 5K, $I_c^{dep}$ ~21 μA. We note that the $I_c$ of the long wires is almost half of the calculated $I_c^{dep}$ as previously observed[28,30]. Our results therefore suggest that the inhomogeneity is the main reason for the low observed critical currents in long meanders.

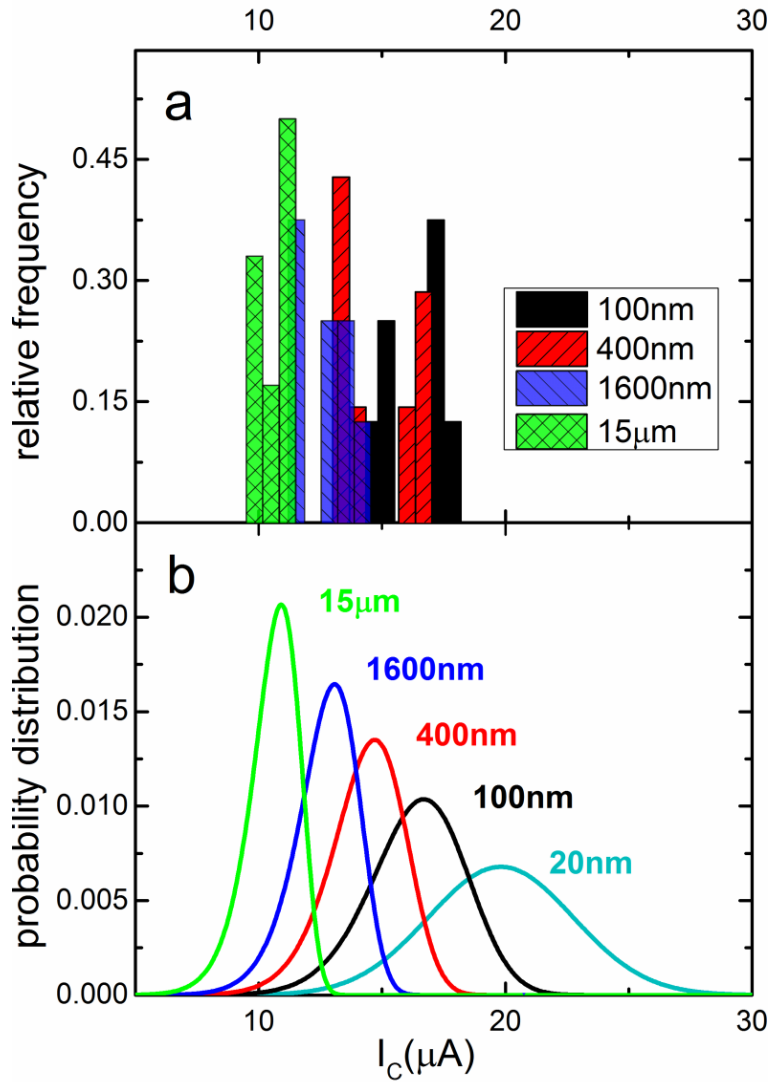

Figure 3. The experimental $I_c$ histograms (panel a) are compared to the $I_c$ probability distribution calculated from the model (panel b). The light blue curve in panel b is the calculated probability distribution for the 20nm section.

We developed a simple microscopic model to endorse our experimental observations. The nanowire is modelled as a chain of identical sections (width 100nm and length s) each of which has a different critical current $I_{ci}$. We assume that the probability for a section to have a certain $I_{ci}$ value is given by a Gaussian probability distribution with mean value μ and standard deviation σ. For a wire of N sections (N=L/s) the critical current $I_{c,wire}$ will be given by the minimum $I_{ci}$ among the N different values. For each wire length and for a given (μ, σ), we calculate the probability that the entire chain of sections has the critical current $i_{ci}$, where 0 <$i_{ci}$ < 50μA. The model starts with an initial guess for (μ, σ) and proceeds by iterating the probability calculation for different (μ, σ). The model stops when the pair (μ, σ) minimizing the $\chi^2$ between the modeled and experimental $\bar{I}_c$ is found, assuming a section length s=20nm. The best agreement for the four wire lengths is found for μ=19.82μA and σ=2.94μA. The calculated $I_c$ probability distributions (for $I_c$ intervals of 0.2μA) are reported in figure 3 (panel b) together with the Gaussian distribution for the $I_c$ of the 20nm section. This graph, when compared to the histograms of panel a, clearly shows that the decrease in both $\bar{I}_c$ and σ is reproduced quantitatively by this simple model. In addition, in figure 2 the average $I_c$ values and the standard deviations predicted by the model are superimposed to the experimental results. The agreement further proves that with simple assumptions we can quantitatively describe the electrical behavior of our devices. We note that a comparable agreement can be found with any section length s≤100nm, so that the present data does not allow a more precise estimation of the correlation length. It should be noted that the best estimate of the average critical current for the 20nm section is very close to the theoretical depairing current, which further supports our interpretation. In our search for the cause of this inhomogeneity, we extensively investigated SEM micrographs of our nanowires. Since the width, measured with the image processing software WSxM[32], has a standard deviation of 2.7%, smaller than the relative variation in the $I_c$: $\Delta I_c/\bar{I}_c$~6.1%, we tentatively attribute the inhomogeneity to the variation of film thickness and/or crystal properties.

Since a long wire can only be biased at almost half of the $\bar{I}_c$ of the shortest wire, we expect the internal quantum efficiency (QE) of SSPDs to be strongly limited. To extract more information about the effect of inhomogeneities on the efficiency, we studied the optical response of selected wires. For each wire length, we selected 5-6 devices and measured them during two consecutive cooling runs. To ensure temperature reproducibility, we made sure that the measured $I_c$ values were the same within 1.2% in both cooling runs. During the optical characterization, a laser beam at λ=1300nm is focused onto a spot with full-width-half-maximum of 20 μm to illuminate one detector at a time. With respect to the circuit used during the electrical characterization, the 50Ωcap is removed and a 4dB attenuator connected to a series of four amplifiers (with 60dB total amplification) is now added to the RF port of the bias-T. The detector output signal coming from the amplifiers is then sent to a counter.

For each nanowire length, the light power is chosen in order to avoid multi-photon response and device heating. For this last purpose, the ratio between the device critical current measured with and without laser spot is always kept above 0.96. In these conditions, we recorded detector counts with (photon counts, PC) and without (dark counts, DC) laser spot while varying the bias current, $I_b$. The QE is calculated by normalizing the difference between PC and DC by the number of photons impinging on the detector active area (100nm x L). We note that this definition of QE slightly differs from the device QE used for meander detectors where the entire area of the device is considered.

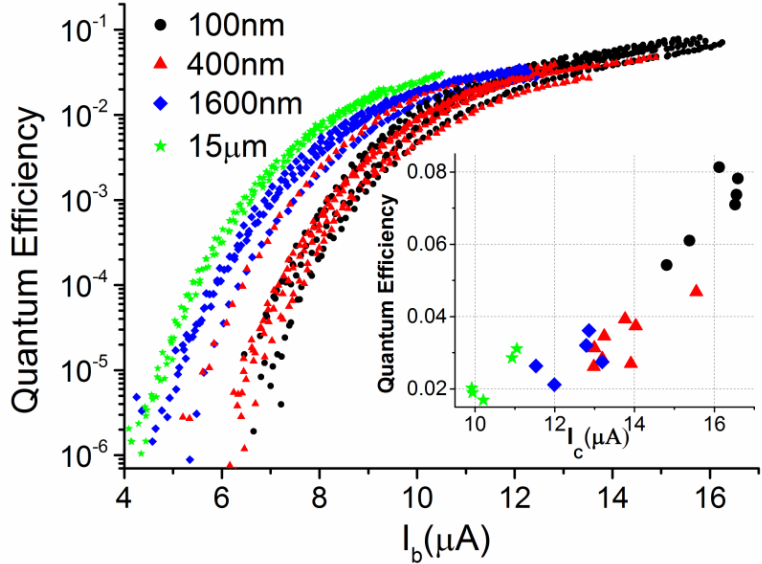

Figure 4. Quantum efficiency curves for devices of different lengths are plotted as function of $I_b$. Inset: maximum quantum efficiency data is plotted as function of device $I_c$.

The QE curves in figure 4 (main panel) show that a detector responds to incoming light with its highest efficiency only when it is biased very close to its $I_c$. As a consequence of the $I_c$ decrease with increasing detector length, the QE curves of short nanowires always extend to higher $I_b$ and QE values with respect to the long nanowire case. The previous observation is made explicit by the inset of figure 4 where the maximum quantum efficiency values are plotted as a function of the $I_c$. The inset clearly shows how a lower $I_c$ corresponds to a lower QE[31]. In addition, the curves of the main panel show that at fixed and low $I_b$ the long wires appear to be more efficient than the short ones. This is related to the fact that some sections of the long wires operate at a bias current closer to their critical current. However, when averaging over many devices, we would expect the same efficiency in long and short wires, since long wires effectively average over many short wires. It could result from the limited statistics available, whereby short wires with low $I_c$, and therefore much more efficient at low $I_b$, are not likely to be measured.

In summary, we have investigated the inhomogeneous nature of NbN nanowires by systematically studying the critical current and efficiency of wires with different lengths. The observed trends cannot be explained by the presence of isolated constrictions, but rather strongly indicate that the wires are inhomogeneous at a length scale shorter than or equal to 100nm. While more investigations are needed in order to determine the physical origin of the non-uniformity of the critical current, small fluctuations in the NbN film thickness being a likely candidate. We note that the proposed experimental method, based on the measurement of the critical current for a number of short wires, enables a simple and reproducible assessment of the film homogeneity and thereby provides more information than extensive QE measurements on long meanders.

The authors would like to thank Saeedeh Jahanmirinejad and Giulia Enrica Digeronimo for technical assistance, Sartoon Fattah poor for support with MATLAB code and Jelmer Renema, Michiel de Dood and Martin van Exter for useful discussions. The nanofabrication was carried out in the Nanolab@TU/e cleanroom facility. This work is part of the research programme of the Foundation for Fundamental Research on Matter (FOM), which is financially supported by the Netherlands Organisation for Scientific Research (NWO), is also supported by NanoNextNL, a micro- and nanotechnology program of the Dutch Ministry of Economic Affairs, Agriculture and Innovation (EL$\& $I) and 130 partners and by the Dutch Technology Foundation STW, applied science division of NWO, the Technology Program of the Ministry of Economic Affairs.

[1]  G. Goltsman, O.Okunev, G. Chulkova, A. Lipatov, A. Semenov, K.Smirnov, B. Voronov, A. Dzardanov, C. Williams, R. Sobolewski, Appl. Phys. Lett **79**,705, (2001)

[2]  C.M. Natarajan, M.G. Tanner and R.H. Hadfield, Supercond. Sci. Technol. **25**, 063001, (2012).

[3]  A.Pearlman, A. Cross, W. Słysz, J. Zhang, A. Verevkin, M. Currie, A. Korneev, P. Kouminov, K. Smirnov, B. Voronov, G. Gol'tsman, and Roman Sobolewski, IEEE Trans. Appl. Supercond. **15**, 579, (2005).

[4] T.Yamashita, S.Miki, K.Makise, W.Qiu, H.Terai, M.Fujiwara, M.Sasaki and Z.Wang, Appl. Phys. Lett. **99**, 161105, (2011)

[5]  B.Baek, A.E.Lita, V.Verma and S.W.Nam, Appl. Phys. Lett. **98**, 251105 (2011).

[6]  V.B. Verma, F. Marsili, S. Harrington, A.E. Lita, R.P. Mirin and S.W. Nam, Appl. Phys. Lett. **101**, 251114 (2012).

[7]  F. Marsili, V.B.Verma, J.A.Stern, S.Harrington, A.E.Lita, T.Gerrits, I.Vayshenker, B.Baek, M. D. Shaw, R.P.Mirin and S.W.Nam, Nature Phot., **7**, 210, (2013)

[8] V.Verma, A.Lita, M.R. Vissers, F.Marsili, D.P. Pappas, R.P.Mirin, and Sae Woo Nam, Appl.Phys Lett. **105**, 022602 (2014).

[9] V.B.Verma, B.Korzh, F.Bussières, R.D.Horansky, A.E.Lita, F.Marsili, M.D.Shaw, H.Zbinden, R.P.Mirin and S.W.Nam, Appl. Phys. Lett. **105**, 122601 (2014).

[10]  H.Takesue, S.W.Nam, Q.Zhang, R.H.Hadfield, T.Honjo, K.Tamaki and Y.Yamamoto, Nature Photonics **1**, 343 (2007).

[11]  D. Stucki, C. Barreiro, S. Fasel, J. Gautier, O. Gay, N. Gisin, R. Thew, Y. Thoma, P. Trinkler, F. Vannel and H. Zbinden, Opt. Express **17**, 13326, (2009)

[12]  D. Bitauld, F. Marsili, A. Gaggero, F. Mattioli,R. Leoni, S. Jahanmiri Nejad, F. Le´vy, and A. Fiore, Nano Lett. **10**, 2977, (2010)

[13] C. Zinoni, B. Alloing, L.H. Li, F. Marsili, A. Fiore, L. Lunghi, A. Gerardino, Yu B. Vakhtomin, K.V. Smirnov, G.N. Gol'tsman, Appl. Phys. Lett. **91**, 031106, (2007).

[14] M.Allman, V.B.Verma, R.Horansky, F.Marsili, J.A.Stern, M.D.Shaw, A.D.Beyer, R.P.Mirin and S.W.Nam, CLEO: Applications and Technology, AW3P.3 (2014)

[15] A. Divochiy, F. Marsili, D. Bitauld, A. Gaggero, R. Leoni, F. Mattioli, A. Korneev, V. Seleznev, N. Kaurova, O. Minaeva, G. Gol'Tsman, K. G. Lagoudakis, M. Benkhaoul, F. Levy and A. Fiore, Nature Phot. **2**, 302 (2008).

[16] Z. Zhou, S. Jahanmirinejad, F. Mattioli, D. Sahin, G. Frucci, A. Gaggero, R. Leoni and A. Fiore, Opt. Express **22**, 3475-3489 (2014).

[17] V.B.Verma, R.Horansky, F. Marsili, J.A.Stern, M.D.Shaw, A.E.Lita, R.P.Mirin and S.W.Nam, Appl. Phys. Lett. **104**, 051115 (2014)

[18] M.Shaw, K.Birnbaum, M.Cheng, M.Srinivasan, K.Quirk, J.Kovalik, A.Biswas, A.D. Beyer, F.Marsili, V.Verma, R.P.Mirin, S.W.Nam, J.A.Stern and W.H.Farr, CLEO: Science and Innovations, SM4J. 2 (2014)

[19] J. J. Renema, R. Gaudio, Q. Wang, Z. Zhou, A. Gaggero, F. Mattioli, R. Leoni, D. Sahin, M. J. A. de Dood, A. Fiore and M. P. van Exter, Phys. Rev. Lett. **112**, 117604 (2014)

[20] A.J. Kerman, E.A. Dauler, J.K.W. Yang, K.M. Rosfjord, V. Anant, K.K. Berggren, G. N. Goltsman and B.Voronov, Appl. Phys. Lett. **90**, 101110, (2007)

[21] R.H. Hadfield, P.A. Dalgarno, J.A. O'Connor, E.Ramsay, R.J. Warburton, E.J. Gansen, B. Baek, M.J. Stevens, R.P. Mirin and S.W. Nam, Appl. Phys. Lett. **91**,241108, (2007)

[22] A. Gaggero, S. Jahanmiri Nejad, F. Marsili, F. Mattioli, R. Leoni, D. Bitauld, D. Sahin, G. J. Hamhuis, R. Nötzel, R. Sanjines and A. Fiore, Appl. Phys. Lett. **97**, 151108 (2010).

[23] J.P. Sprengers, A. Gaggero, D. Sahin, S. Jahanmiri Nejad, F. Mattioli, R. Leoni, J. Beetz, M. Lermer, M. Kamp, S. Höfling, R. Sanjines, A. Fiore, Appl. Phys. Lett. **99**, 181110 (2011).

[24] D. Sahin, A. Gaggero, Z. Zhou, S. Jahanmiri Nejad, F. Mattioli, R. Leoni, J. Beetz, M. Lermer, M. Kamp, S. Höfling, A. Fiore, Appl. Phys. Lett. **103**, 111116, 2013.

[25] A.J. Kerman, E.A. Dauler, W.E. Keicher, J.K.W. Yang, K.K. Berggren, G. N. Goltsman and B.Voronov, Appl. Phys. Lett. **88**, 111116, (2006)

[26] F. Marsili, D. Bitauld, A. Gaggero, S. Jahanmirinejad, R. Leoni, F. Mattioli and A. Fiore, New J. Phys. **11**, 045022, (2009)

[27] J. Bardeen, Rev. Mod. Phys. **34**, 667 (1962).

[28] D. Henrich, P. Reichensperger, M. Hofherr, J. M. Meckbach, K. Il'in, M. Siegel, A. Semenov, A. Zotova, and D. Yu. Vodolazov, Phys. Rev. B **86**, 144504, (2012).

[29] A. Semenov, B. Günther, U. Böttger, H.W. Hübers, H. Bartolf, A. Engel, A. Schilling, K. Ilin and M. Siegel, R. Schneider, D. Gerthsen and N. A. Gippius, Phys. Rev. B **80**, 054510, (2009).

[30] M. Hofherr, D. Rall, K. Ilin, M. Siegel, A. Semenov, H.W. Hübers and N.A. Gippius, J. Appl. Phys. **108** 014507 (2010).

[31] We note that there is an experimental uncertainty of 20-30% in the QE data due to possible variation of the polarization state.

[32]  I.Horcas, R. Fernandez, J.M. Gomez-Rodriguez, J. Colchero, J. Gomez-Herrero and A. M. Baro, Rev. Sci. Instrum. **78**, 013705, (2007).